\documentclass[pdflatex,sn-mathphys-num]{sn-jnl}

\usepackage{graphicx}%
\usepackage{multirow}%
\usepackage{amsmath,amssymb,amsfonts}%
\usepackage{amsthm}%
\usepackage[title]{appendix}%
\usepackage{xcolor}%
\usepackage{textcomp}%
\usepackage{manyfoot}%
\usepackage{booktabs}%
\usepackage{algorithm}%
\usepackage{algorithmicx}%
\usepackage{algpseudocode}%
\usepackage{listings}%
\usepackage{microtype}%

\theoremstyle{thmstyleone}%
\theoremstyle{thmstyletwo}%

\theoremstyle{thmstylethree}%

\hypersetup{bookmarksdepth=subsubsection,hypertexnames=false}
\begin{document}
\title[COACH]{COACH: Toward the performance limit of traditional hybrid density functionals for molecular chemistry}

\author*[1]{\fnm{Jiashu} \sur{Liang}}\email{jsliang@berkeley.edu}

\author*[1,2]{\fnm{Martin} \sur{Head-Gordon}}\email{mhg@cchem.berkeley.edu}

\affil*[1]{\orgdiv{Kenneth S. Pitzer Center for Theoretical Chemistry, Department of Chemistry}, \orgname{University of California at Berkeley}, \orgaddress{\city{Berkeley}, \state{CA}, \postcode{94720}, \country{USA}}}

\affil[2]{\orgdiv{Chemical Sciences Division}, \orgname{Lawrence Berkeley National Laboratory}, \orgaddress{\city{Berkeley}, \state{CA}, \postcode{94720}, \country{USA}}}

\abstract{Density functional theory (DFT) offers an exceptional balance between accuracy and efficiency, but practical density functional approximations face an unavoidable trade-off among simplicity, accuracy, and transferability.
Here we present a systematic functional-development protocol that combines constraint enforcement, flexible functional forms, and modern optimization within a predefined application domain and functional class.
Applying this strategy to the range-separated hybrid (RSH) meta-GGA framework, we obtain the carefully optimized and appropriately constrained hybrid (COACH) functional.
Across broad molecular benchmarks, COACH improves both accuracy and transferability relative to leading RSH meta-GGAs, including \(\omega\)B97M-V, while retaining the computational practicality of its rung.
These results suggest that COACH comes closer to the performance limit of traditional hybrid meta-GGA functionals for general molecular ground-state chemistry.
Further substantial progress will likely require richer nonlocal or orbital-dependent information than is available from semilocal density ingredients combined with simple exact-exchange mixing.}

\maketitle
\makeatletter
\let\@oddhead\@empty
\let\@evenhead\@empty
\makeatother

\section{Introduction}\label{sec1}

\begin{figure}[htbp]
\centering
\includegraphics[width=0.7\textwidth]{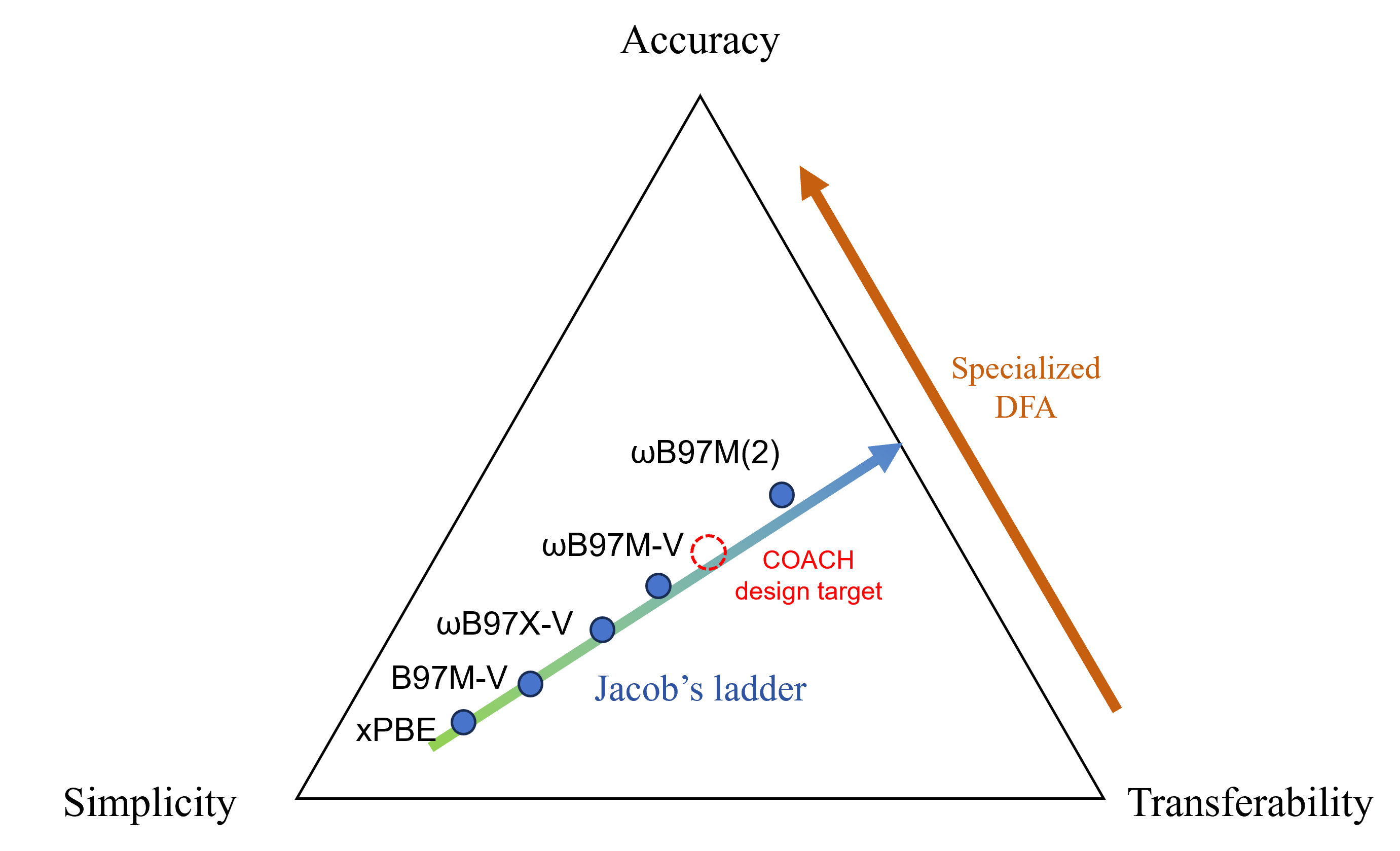}
\caption{Heuristic impossible triangle of simplicity, accuracy, and transferability in density functional approximation development.
The blue points represent the best-performing DFAs on each rung in GSCDB137, and the green-blue arrow indicates the usual improvement in both accuracy and transferability on climbing Jacob's Ladder; the orange arrow represents specialization at a fixed rung, and the dashed red circle represents the qualitative COACH design target.}
\label{fig:impossible_triangle}
\end{figure}

Density functional theory (DFT) is a cornerstone of molecular and materials simulation because it balances computational cost and accuracy \cite{mardirossian2017thirty}.
In principle, the Kohn--Sham formulation provides an exact description of ground-state properties as a functional of the electron density \cite{hohenberg1964inhomogeneous,kohn1965self}.
However, there exists an impossible triangle in density functional approximation (DFA) development (Figure \ref{fig:impossible_triangle} as a heuristic illustration): simplicity, accuracy, and transferability cannot be satisfied simultaneously in practice.
Simplicity corresponds to computational efficiency.
Even if the exact exchange--correlation (XC) functional were available, its complexity would likely limit practical advantages over advanced wavefunction methods.
For example, the exact functional is not fully differentiable, as manifested in the flat-plane conditions \cite{yang2016communication}, which can be viewed as the price of integrating away all but one electron degree of freedom.
Jacob's Ladder~\cite{perdew2005prescription}, introduced in Figure~\ref{fig:impossible_triangle}, provides a useful five-rung hierarchy of DFAs, from the simplest to the most sophisticated, in which higher accuracy and transferability generally come at increased computational cost.
In Figure~\ref{fig:impossible_triangle}, the blue points represent the best-performing DFAs on each rung in GSCDB137~\cite{liang2025gold}.
Within a given rung, however, there will always be a trade-off between transferability and accuracy.
Here, accuracy refers to predictive accuracy within specified domains, whereas transferability refers to the size of the domain over which high accuracy is maintained.
Given a fixed level of simplicity, we define the performance limit operationally as the Pareto frontier on which neither accuracy within the specified domains nor the size of the high-accuracy domain can be improved without degrading the other.
Here, we apply a systematic functional-development protocol to move toward the performance limit of hybrid meta-GGA functionals and obtain the carefully optimized and appropriately constrained hybrid (COACH) functional.
In the protocol, we specify the target molecular domains before optimization, thereby setting the intended transferability; functional development then focuses on improving accuracy within those domains.

To demonstrate our protocol, we choose the hybrid meta-generalized-gradient approximations (hmGGA) form (rung 4) as the target level of simplicity for the following reasons: First, it is the most widely used functional form in chemistry, as present-day double hybrids (DH, rung 5) suffer from more practical problems \cite{martin2020empirical,liu2025revisiting,liang2025gold}.
Second, Jacob's Ladder holds in most cases (the best higher-rung functionals outperform those below), except that hmGGAs have not consistently surpassed hybrid GGAs for electric field properties, excitation energies, and vibrational frequencies \cite{liang2025gold,liang2022revisiting}.
Given the vast functional space available for hmGGA\cite{Mardirossian:2016}, it remains possible to discover more transferable and accurate functionals within this rung.
For transferability, long-range Hartree-Fock exchange complicates plane-wave implementations and can introduce pathologies at some metallic Fermi levels; therefore, we do not target periodic metals or plane-wave implementations.
Moreover, magnetic response properties require current-DFT for a rigorous treatment, and excitation properties involve new errors from the adiabatic approximation, both of which are beyond the present scope.
Accordingly, this work focuses on transferability across molecular ground-state properties evaluated with localized basis sets.
Within this application domain, we focus more specifically on traditional B97-like RSH meta-GGA functionals that combine polynomial expansions in bounded density-gradient and kinetic-energy-density variables with standard error-function range separation and simple fixed exchange mixing.

There are two complementary philosophies guiding DFA development.
Non-empirical functionals enforce exact constraints and recover known limits with minimal reliance on empirical data.
For example, the mGGA SCAN \cite{sun2015strongly} satisfies 17 exact constraints and achieves good accuracy for both molecular and solid-state systems.
The strength of non-empirical functionals lies in their reliability and broad transferability, as they are grounded in fundamental physical principles.
However, their constraint-based design can limit flexibility and thus accuracy in chemical applications \cite{mardirossian2017thirty}.
In contrast, semi-empirical functionals, pioneered by Becke \cite{Becke1993,Becke1993Density,becke1997density}, introduce flexible expansions with coefficients trained to high-quality data.
This approach has produced many leading DFAs across rungs, including B97M-V \cite{Mardirossian:2015}, M06-L \cite{zhao2006new}, $\omega$B97X-V \cite{Mardirossian:2014}, $\omega$B97M-V \cite{Mardirossian:2016}, and CF22D \cite{liu2023supervised}, which excel within their training domains.
Among them, $\omega$B97M-V \cite{Mardirossian:2016} is today's most accurate hybrid (rung 4) functional by most measures \cite{liang2025gold}.
However, functionals with similar simplicity and in-domain accuracy can still differ in transferability owing to differences in the quality and coverage of their training protocols; for example, some functionals that are highly accurate for thermochemistry have been shown to yield less accurate electron densities \cite{medvedev2017density}.
Indeed, $\omega$B97M-V (and other existing semi-empirical functionals) can underperform when systems deviate substantially from their training data, as demonstrated in our recent GSCDB137 benchmark \cite{liang2025gold}.

These observations motivate a synthesis of the two philosophies: enforce constraints and select physically meaningful variables to ensure transferability, while adopting flexible functional forms and modern optimization techniques to achieve high accuracy.
Our protocol has four pillars: train and validate on a large, diverse, and balanced database (we employ Ref. \citenum{liang2025gold}) to discourage overfitting; adopt a highly flexible but structured functional form to explore a broad region of functional space; enforce exact constraints analytically where possible and numerically where necessary; and employ best-subset selection algorithms to improve accuracy without serious overfitting.
These strategies are detailed in the Supplementary Information and resulted in our carefully optimized and appropriately constrained hybrid (COACH) functional.
Built on these four pillars, our protocol systematically examines all principal functional-form choices predefined for the claimed class.
In the following sections, we show that COACH substantially improves accuracy and transferability relative to leading range-separated hybrid (RSH) mGGAs, particularly $\omega$B97M-V, across a broad range of molecular benchmarks, supporting the hypothesis that COACH may approach the performance limit of this class.

\section{Results}\label{sec2}

\subsection{Design of the COACH functional}

COACH is a RSH \cite{gill1996coulomb,leininger1997combining,iikura2001long} mGGA with B97-type inhomogeneity corrections, where the only approximations are contained in the exchange ($x$) correlation ($c$) energy.
The COACH exchange ($x$) energy is constructed from semi-local (SL), and exact (HF) contributions, where the latter is range-separated into short-range (sr) and long-range (lr) contributions,
\begin{equation}
    E_x = E_{x,\mathrm{sr}}^{\mathrm{SL}} + c_{x,\mathrm{sr}}^{\mathrm{HF}} E_{x,\mathrm{sr}}^{\mathrm{HF}} + c_{x,\mathrm{lr}}^{\mathrm{HF}} E_{x,\mathrm{lr}}^{\mathrm{HF}},
\end{equation}
with $c_{x,\mathrm{lr}}^{\mathrm{HF}}=1$ (long-range corrected).
The correlation energy is evaluated from SL same-spin (ss) and opposite-spin (os) contributions, and a non-local term describing the dispersion (disp) energy \cite{vydrov2010nonlocal,caldeweyher2019generally},
\begin{equation}
E_c = E_{c,\mathrm{ss}}^{\mathrm{SL}} + E_{c,\mathrm{os}}^{\mathrm{SL}} + E_c^{\mathrm{disp}}.
\end{equation}
Each of the semi-local $xc$ terms take the form of
\begin{equation}
E_{x(c)}^{\mathrm{SL}}=\int e_{x(c)}^{\mathrm{base}}\,F_{\mathrm{corr}}\,g_{x(c)}\,d\mathbf r,
\end{equation}
where $e_{x(c)}^{\mathrm{base}}$ is a uniform-electron-gas (UEG) based energy density and $F_{\mathrm{corr}}$ is a correction factor that enforces physical constraints.
On the other hand, $g_{x(c)}$ is an additional correction factor that will be designed using training and validation data, whose general form is a 2-dimensional expansions in bounded, dimensionless variables, $u$ and $v$
\begin{equation}
g_{x\sigma}=\sum_{i=0}^{M}\sum_{j=0}^{N} c_{x,ij} f_i(u) f_j(v).
\end{equation}
$u$ and $v$ depend on the electron density gradient and kinetic energy density, respectively, $f_i(u)$ is an $i^\mathrm{th}$ order term (e.g. $u^i$ for simple polynomials) and $c_{x,ij}$ are linear coefficients to be appropriately determined.
As limits for the expansions, we set $M=11$ and $N=7$ (289 linear parameters total including $c_{x,\mathrm{sr}}^{\mathrm{HF}}$).

Our optimization protocol is briefly summarized in the Methods and fully elaborated in the SI.
In accordance with our design philosophy, this protocol systematically covers the key ingredients of functional construction, including the choice of base forms, nonlinear parameters, dimensionless variables, and polynomial expansion bases, together with the strategy used to design experiments and determine the final selections.
The protocol is intended to be general rather than specific to COACH alone and can in principle be used to train high-performing mGGAs, hybrid functionals, and double hybrids for a chosen target domain.
In this work, COACH provides a competitive, constraint-aware realization of this philosophy and protocol within the predefined functional class.

The final COACH functional is the design point that satisfies the chosen constraints and numerical-stability targets on the optimization meshes while achieving the lowest overall mean error.
It contains 73 optimized linear parameters and 3 optimized nonlinear parameters, all fully documented in the SI.
In the following subsections, we compare the constraints satisfied by COACH and the level of error achievable for chemical energy differences and molecular properties against other ``best-of-breed'' hybrid functionals.

\subsection{Constraints}

The design of non-empirical density functionals is guided as far as possible by physical constraints.
In the design of COACH, we examined which of the 17 commonly cited mGGA exact constraints\cite{sun2015strongly} can be satisfied within a RSH mGGA framework without degrading the quality of chemical predictions.
While a detailed analysis is given in the SI, Table~\ref{tab:constraints} compares the constraint satisfaction of SCAN, COACH, \(\omega\)B97M-V \cite{Mardirossian:2016} (the parent functional form), and CF22D \cite{liu2023supervised}.
Relative to \(\omega\)B97M-V, we were able to increase the number of fully satisfied constraints in COACH by incorporating several physically motivated design choices.
Specifically, COACH satisfies X05 by adopting SCAN’s non-uniform scaling correction for exchange and satisfies C09 and C12 by using SCAN correlation (\(\alpha=1\)) as the base correlation functional.
The final coefficients certify the semi-local X06 upper bound, supporting its classification as partially satisfied, while the analytic $2\beta$ factor gives the one-electron limit for C11.
The kinetic-energy-density indicator \(\beta_\sigma=(\tau_\sigma-\tau_\sigma^W)/(\tau_\sigma+\tau_\sigma^{\mathrm{UEG}})\) distinguishes single-orbital and slowly varying regions while avoiding some numerical problems associated with \(\alpha\)~\cite{furness2019enhancing}.
These additions, together with the constraint-aware training protocol described in the SI, make COACH a semi-empirical functional that satisfies more physical constraints than any existing semi-empirical hybrid functional.
Importantly, improved constraint satisfaction does not by itself guarantee superior accuracy.
The limited form of approximate functionals mean that there may be a tradeoff between constraint satisfaction and the level of chemical accuracy.
Rather, satisfaction of exact constraints is expected to reduce unphysical behavior and make density functional approximations more predictive over the immense space of many-electron systems, consistent with the philosophy underlying non-empirical functional development~\cite{kaplan2023predictive}.
In this sense, COACH is expected to be transferable across chemically diverse applications, even though COACH retains a much more  highly flexible parameterization (73 linear parameters) relative to \(\omega\)B97M-V (12 linear parameters).

\begin{table}[htbp]
\caption{Exact constraints and satisfaction status for SCAN, COACH, $\omega$B97M-V, and CF22D. Here, Y indicates fully satisfied, P indicates partially satisfied, and a blank entry indicates not satisfied.}
\label{tab:constraints}
\centering
\begin{tabular}{lcccc}
\toprule
Condition & SCAN & COACH & $\omega$B97M-V & CF22D \\
\midrule
total number fully satisfied & 17 & 7 & 4 & 3 \\
total number fully + partially satisfied & 17 & 12 & 9 & 5 \\
\midrule
X01: negativity & Y & P & Y &  \\
X02: spin-scaling & Y & Y & Y & Y \\
X03: uniform density scaling & Y & P & P &  \\
X04: 4th order gradient expansion & Y & P & P &  \\
X05: non-uniform density scaling & Y & Y & P & P \\
X06: tight bound for spin-unpolarized 2e densities & Y & P &  &  \\
C07: non-positivity & Y &  &  &  \\
C08: 2nd order gradient expansion & Y &  &  &  \\
C09: uniform scaling (high density limit) & Y & Y &  &  \\
C10: uniform scaling (low density limit) & Y & Y & Y & Y \\
C11: zero correlation for 1e densities & Y & Y &  &  \\
C12: non-uniform density scaling & Y & Y & P & P \\
XC13: size-extensivity & Y & Y & Y &  Y \\
XC14: general Lieb-Oxford bound & Y & P & P &  \\
XC15: no low-density spin-polarization dependence & Y &  &  &  \\
XC16: static linear response of UEG & Y &  &  &  \\
XC17: Lieb-Oxford bound for 2e densities & Y &  &  &  \\
\bottomrule
\end{tabular}
\end{table}

\subsection{Performance on Diverse Chemical Systems}

\begin{figure}[htbp]
    \centering
    \includegraphics[width=\textwidth]{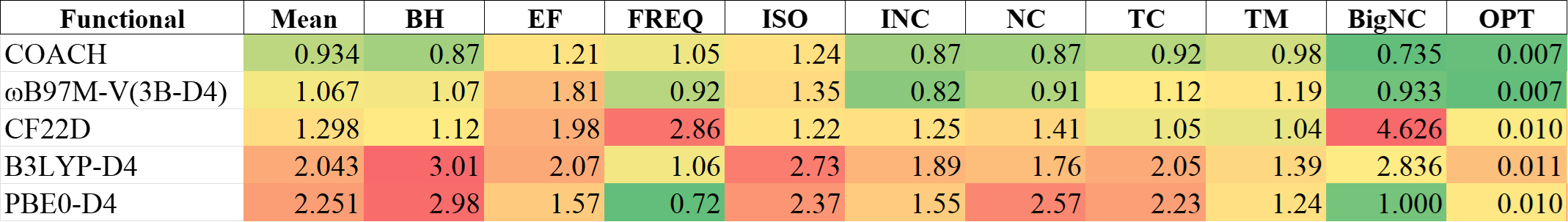}
    \caption{Category-resolved performance of COACH and comparison functionals. Values are normalized error ratios (NERs) for BH, EF, FREQ, ISO, INC, NC, TC, TM, BigNC, and the overall mean; lower values indicate better performance. For each GSCDB137 data set, the NER is the applicable data-set-specific error metric divided by the average of the second-, third-, and fourth-lowest errors among the 14 evaluated hybrid functionals, following Ref.~\citenum{liang2025gold}. BigNC is evaluated on the L14 and vL11 large-complex benchmarks \cite{Lao2024CanonicalCCBindingBenchmark} using PBE0-D4 MAEs as standard errors (L14: 0.68~kcal/mol; vL11: 1.05~kcal/mol). The OPT column reports the root mean square of per-molecule RMSDs between optimized and reference geometries in units of Angstroms (W4-11-GEOM and SE sets).}
    \label{fig:performance_heatmap}
\end{figure}

Figure~\ref{fig:performance_heatmap} summarizes category-resolved performance using normalized error ratios (NERs) for BH (reaction barrier heights), EF (responses to applied electric fields), FREQ (vibrational frequencies, which probe potential-energy-surface curvature), ISO (relative energies of isomers), INC and NC (intra- and intermolecular noncovalent interactions), TC (main-group thermochemical energy differences), TM (transition-metal chemistry), BigNC (large noncovalent complexes), and the overall mean.
Detailed descriptions of the individual GSCDB137 data sets are provided in Ref.~\citenum{liang2025gold}.
The figure also reports geometry-optimization performance (OPT) as the root mean square (RMS) of the per-molecule RMS deviations between optimized and reference geometries in units of Angstroms.
For each data set, the NER is its data-set-specific error metric divided by the corresponding standard error, following the GSCDB137 definition \cite{liang2025gold}.
The overall mean NER is the arithmetic mean of these data-set-level NERs over the included GSCDB137 data sets.
Because normalization by a relative denominator can mask uniformly large absolute errors within a data set, the supporting data file reports \(N\), MSE, MAE, RMSE, the data-set-specific error metric, its denominator, and NER for every data set and functional.
BigNC and OPT are not parts of GSCDB137.
For BigNC, we use the L14 and vL11 large-complex benchmarks \cite{Lao2024CanonicalCCBindingBenchmark} and define the standard errors as the mean absolute errors (MAEs) of PBE0-D4 on each set (L14: 0.68~kcal/mol; vL11: 1.05~kcal/mol).
The OPT category includes the W4-11-GEOM~\cite{karton2021evaluation} and SE~\cite{lazzari2025molecular} sets.
The XYZ files, reference data, and evaluation metrics for these sets are available at \url{https://github.com/JiashuLiang/GSCDB}.

\begin{figure}[htbp]
    \centering
    \includegraphics[width=0.7\textwidth]{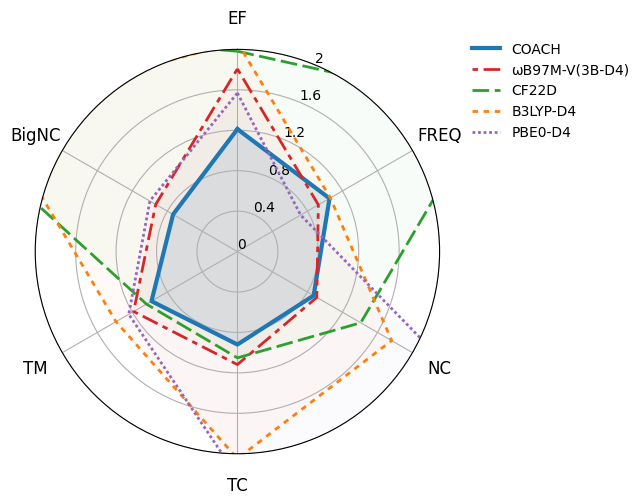}
    \caption{Radar comparison of selected category-level NERs (EF, FREQ, NC, TC, TM, BigNC) for COACH and representative functionals. Smaller radii indicate lower NERs and better performance.}
    \label{fig:performance_radar}
\end{figure}

We compare COACH with two representative modern functionals, \(\omega\)B97M-V and CF22D, and with two widely used traditional hybrids, PBE0 and B3LYP.
To describe noncovalent interactions in larger systems (BigNC), we apply D4 dispersion corrections to PBE0 and B3LYP, and we include the D4-ATM three-body term for \(\omega\)B97M-V using the parameters of \(\omega\)B97M-D4 (this is essential for good results on BigNC).
The only exception is CF22D, which uses its own fixed dispersion counterpart.
Figure~\ref{fig:performance_radar} provides a compact visualization for selected categories.

COACH achieves the lowest overall mean NER (0.93), representing a 13\% improvement over $\omega$B97M-V (1.07), and it substantially outperforms CF22D, B3LYP-D4, and PBE0-D4.
Notably, COACH is the only tested functional with an overall mean NER below one, suggesting that it comes closer than the other existing hybrid functionals to the performance limit of hybrid functionals.

The traditional hybrids B3LYP-D4 and PBE0-D4 perform poorly for TC and NC relative to \(\omega\)B97M-V and CF22D, where modern functionals have been more heavily trained, whereas their relative underperformance is less pronounced for TM, where training is also more limited.
At the same time, B3LYP-D4 and PBE0-D4 can perform comparably or even better for EF and FREQ, underscoring the need for careful training and validation of new semi-empirical functionals.

Relative to \(\omega\)B97M-V(3B-D4) and CF22D, the largest gains of COACH occur for EF and BigNC, and COACH also leads in BH, ISO, NC, TC, and TM.
On BigNC, the NERs of COACH and \(\omega\)B97M-V decrease from 2.81 and 3.32 without D4-ATM to 0.73 and 0.93 with their respective D4-ATM treatments, showing that three-body dispersion is the main source of the improvement, while the lower no-ATM NER of COACH shows that the underlying functional also contributes.
FREQ is the only category in which COACH is not best-in-class, although the difference is small. This result should be interpreted in the context of the limited frequency data used during development: V30 is the sole FREQ set, was not included in the final coefficient fitting, and contains only vibrational frequencies of small molecular dimers.
For geometry optimization, COACH and \(\omega\)B97M-V perform best, although errors remain low for all functionals.

We note that all data sets shown above, except for OPT, were used either for training or for functional design selection.
Performance on these development data therefore cannot by itself establish out-of-sample transferability.
To obtain a broader post-development assessment, we evaluated COACH on nine additional test sets spanning BH, ISO, INC, NC, TC, TM, and BigNC categories~\cite{galleni2025gw,karton2025highly,hoffman2026h2x100,karton2026isoc7,dahl2026taeptcomp,hughes2026trip50,karton2026w1sn2bh,knattrup2025atmospheric218,lao2026vl27}.
The post-development evaluation protocol is described in the SI.
\begin{table*}[htbp]
\begingroup
\caption{Post-development evaluation of COACH on additional benchmark sets, reported as MAE/RMSE in kcal mol$^{-1}$; H2X100 values are normalized per pairwise interaction and TAE-PTComp values are scaled per atom.}
\label{tab:external_validation}
\centering
\small
\resizebox{\textwidth}{!}{%
\begin{tabular}{llrccc}
\toprule
Benchmark & Target quantity & $N$ & COACH & Comparator & Comparator result \\
\midrule
3dCIPEA & IP/EA & 92 & 6.85/11.23 & $\omega$B97M-V$^{a}$ & 7.06/10.90 \\
GDB9-W1-F12 & Atomization energy & 3366 & 1.25/1.55 & M06-2X$^{b}$ & 1.84/2.37 \\
H2X100 & Binding energy/pair & 100 & 0.079/0.092 & CAM-B3LYP-D4$^{b}$ & --/0.08 \\
ISOC7 & Isomerization energy & 1308 & 1.00/1.24 & $\omega$B97M-D4$^{b}$ & 1.38/1.62 \\
TAE-PTComp & Atom-scaled TAE & 2097 & 0.74/1.06 & $\omega$B97M-V$^{b}$ & 0.8/1.3 \\
TRIP50 kinetics & Triplet barriers & 100 & 1.38/1.95 & $\omega$M06-D3$^{b}$ & 1.486/-- \\
TRIP50 thermochemistry & Triplet reaction energies & 50 & 1.54/1.86 & $\omega$M06-D3$^{b}$ & 1.470/-- \\
W1-SN2-BH & Anionic S$_\mathrm{N}$2 barriers & 1881 & 1.97/2.54 & $\omega$B97M-V$^{b}$ & 1.84/2.39 \\
atmospheric218 & Cluster binding energy & 218 & 0.47/0.65 & $\omega$B97M-V$^{a}$ & 0.50/0.73 \\
vL27$^{\dagger}$ & Large-complex binding & 27 & 0.94/1.12 & $\omega$B97M-D4$^{b}$ & 1.28/1.52 \\
\bottomrule
\end{tabular}
}
\begin{flushleft}
\footnotesize $^{a}$Recomputed using the same local Q-Chem protocol as COACH.
$^{b}$Best hybrid result reported in the source study; these rows are not uniform same-code comparisons.
$^{\dagger}$Eleven vL27 complexes are also in BigNC/vL11, but vL27 uses new, more accurate references with 7/8 complete-PNO-space extrapolation rather than the vL11 references used during fitting.
\end{flushleft}
\endgroup
\end{table*}

The largest MAE reductions occur for GDB9-W1-F12, vL27, and ISOC7, with reductions of 32\%, 27\%, and 28\%, respectively.
COACH also performs best on TAE-PTComp, TRIP50, and atmospheric218 and remains close to the best-performing functional on 3dCIPEA, H2X100, and W1-SN2-BH, showing that its broad performance extends beyond the development data.

\subsection{Grid and basis convergence}

Grid sensitivity is a major practical challenge for (hybrid) mGGAs.
During COACH development (see SI for details), we adopted the numerically stable dimensionless variable \(\beta\) to avoid the extreme grid sensitivity associated with the \(\alpha\) indicator in SCAN.
We further control \(\gamma_{c,ss}\) to mitigate oscillatory behavior in derivatives along potential-energy curves that can arise in B97-type forms.
Most importantly, we explicitly enforce numerical grid-sensitivity constraints between a large grid (99,590) and an ultra-large grid (250,974) during optimization; here \((99,590)\) denotes 99 radial shells and 590 angular (Lebedev) points per shell.
Details are provided in the SI.

For the final assessment, we evaluated COACH on GSCDB137 using two grid setups, (75,302) and (99,590).
Only seven reactions show an energy difference larger than 0.5 kcal/mol; six of these correspond to atomic energies, and the remaining case is from 3d4dIPSS, which also involves only single atoms.
Therefore, COACH can be safely used with the medium grid (75,302) for most applications, although we recommend (99,590) because our training constraint guarantees grid errors below 0.015 kcal/mol at this resolution.
The only exception is the V30 set, which describes frequencies of molecular dimers.
When the vibrational frequencies are computed by finite differences using (75,302), 115 out of 275 frequencies differ by more than 10~cm\(^{-1}\).
We also tested analytical frequencies, for which 44 out of 331 frequencies differ by more than 10~cm\(^{-1}\); however, a small number of very large deviations (greater than 100~cm\(^{-1}\)) still occur, especially for the lowest frequencies corresponding to the noncovalent interactions.

\begin{figure}[htbp]
    \centering
    \includegraphics[width=0.6\textwidth]{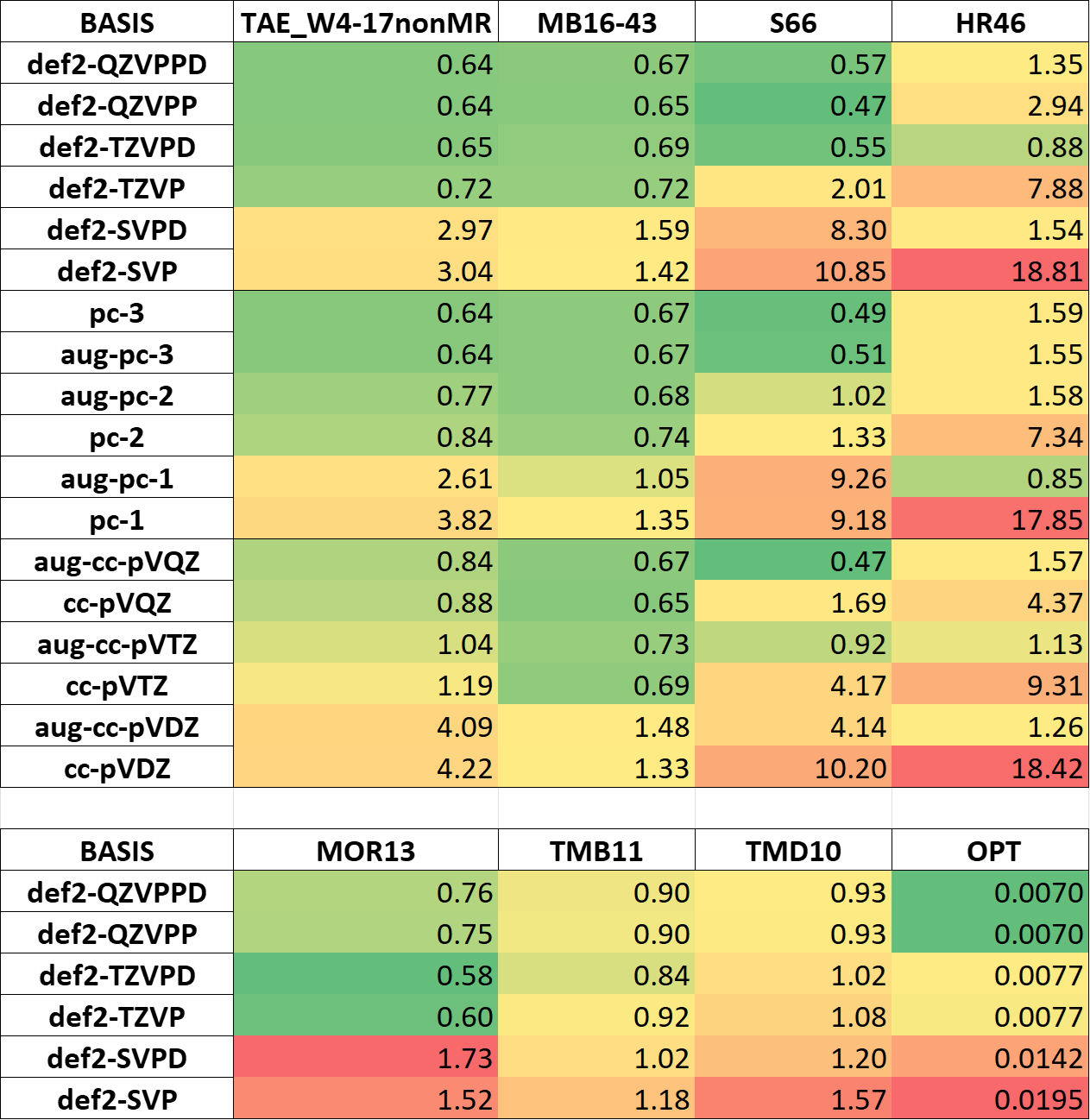}
    \caption{Basis-set convergence of COACH across the Ahlrichs def2, Jensen polarization-consistent (pc), and Dunning correlation-consistent (cc) families. \textbf{Top:} NERs for TAE\_W4-17nonMR, MB16-43, S66, and HR46 using the basis sets def2-QZVPPD, def2-QZVPP, def2-TZVPD, def2-TZVP, def2-SVPD, def2-SVP, pc-3, aug-pc-3, aug-pc-2, pc-2, aug-pc-1, pc-1, aug-cc-pVQZ, cc-pVQZ, aug-cc-pVTZ, cc-pVTZ, aug-cc-pVDZ, and cc-pVDZ. \textbf{Bottom:} NERs for MOR13, TMB11, and TMD10, and performance on the OPT benchmark, using def2-QZVPPD, def2-QZVPP, def2-TZVPD, def2-TZVP, def2-SVPD, and def2-SVP. Except for OPT, all panels report NER values.}
    \label{fig:basis_convergence}
\end{figure}

We also examined basis-set convergence across the Ahlrichs def2, Dunning correlation-consistent (cc), and Jensen polarization-consistent (pc) families \cite{weigend2005balanced,dunning1989gaussian,jensen2002polarization}.
Figure~\ref{fig:basis_convergence} reports NERs for the energy and property benchmarks (TAE\_W4-17nonMR, MB16-43, S66, and HR46) and for the remaining test sets (MOR13, TMB11, and TMD10), as well as an optimization (OPT) benchmark.
For transition-metal sets, only def2 results are shown because the cc and pc families do not cover all required elements.
Overall, COACH performs best with the def2 family.
Notably, def2-TZVPD can yield similar or even better performance than def2-QZVPPD (possibly due to the beneficial error cancellation), making def2-TZVPD an attractive practical choice with substantially reduced computational cost.
We do not recommend double-\(\zeta\) basis sets for energy or property calculations because they produce substantially larger errors, although they may be used for geometry optimization with caution, as the root-mean-squared deviations of geometries remain below 0.02~\AA.
COACH is compatible with the pc family, since (aug)-pc-3 can achieve results comparable to def2-QZVPPD.
In contrast, we do not recommend the cc family: COACH with aug-cc-pVQZ can underperform def2-TZVPD across all tested sets.
This trend is consistent with the design goals of these basis families, with cc optimized primarily for post-HF methods and pc developed specifically for DFT.

\section{Conclusions and Future Direction}\label{sec13}

Guided by the impossible-triangle heuristic, we develop a protocol targeting the best possible balance between accuracy and transferability within a fixed level of simplicity.
For the traditional B97-like RSH meta-GGA class targeted here, we systematically examined all principal functional-form choices predefined in our protocol and used structured optimization to explore a formal combinatorial subset space with a nominal upper-bound count estimated to be \(10^{76}\) times the corresponding count in the original development of \(\omega\)B97M-V, which was widely regarded as the leading existing hybrid functional.
The semilocal components of these meta-GGA functionals are constructed as polynomial expansions in a two-dimensional space of bounded variables derived from the density-gradient norm and kinetic-energy density; increasing the polynomial order provides a systematically convergent route to this chosen functional representation.
Together with COACH's strong performance across broad molecular benchmarks, this systematic design-space exploration supports our hypothesis that COACH may approach the performance limit of the defined class of traditional B97-like RSH meta-GGA functionals.
This remains an empirical, class-specific hypothesis rather than a theoretical error bound.

To facilitate further progress, we also release the complete COACH training protocol.
The protocol, rather than any single parameterization, constitutes the central methodological contribution of this work.
The protocol is not restricted to RSH mGGAs and can be applied to other rungs of Jacob’s Ladder, including mGGAs and double hybrids, as well as to specialized functionals targeting particular applications such as Nuclear Magnetic Resonance properties or solid-state systems.
More broadly, the protocol provides a reusable template for AI-agent-assisted functional development~\cite{duston2026agentic}: once a target domain and the relevant physical ingredients, constraints, and dimensionless variables are defined by researchers or a PLANNER agent, an EXECUTOR agent can use this template to carry out structured functional design, optimization, validation, and auditing.

Further substantial and systematic gains will likely require going beyond semilocal density ingredients combined with simple global or range-separated exact-exchange mixing.
Hybrid density functionals are limited in accuracy by several factors that include the delocalization error (e.g. failure to be exact for 1-electron systems) \cite{bryenton2023delocalization}, and the strong correlation error (fractional spin breakdown when multiple electron configurations are comparably important in the ground state) \cite{cohen2008fractional}.
Higher accuracy for molecules that are not strongly correlated may therefore benefit from richer nonlocal or orbital-dependent information in the exchange--correlation functional.
Promising routes include local hybrids with spatially varying exchange mixing~\cite{maier2019local}, double hybrids with perturbative orbital correlation~\cite{martin2020empirical,liang2025gold}, and explicitly nonlocal or machine-learned nonlocal forms~\cite{luise2025accurate}.
Strong correlation via DFT remains a frontier with promising approaches \cite{gagliardi2017multiconfiguration,su2018describing} and room for new ideas.

\section{Methods}\label{sec:method}

We summarize the key protocol elements here; full derivations, constraint assessments, numerical procedures, and data/weighting tables are provided in the Supplementary Information (SI).

\paragraph{Optimization and constraints.}
We fit the linear coefficients by weighted least squares using a base-functional linearization:
\begin{equation}
L=\frac{1}{2}\left\|\mathbf A\mathbf c-\mathbf b\right\|^2,\quad A_{k,l}=\sqrt{w_k}F_{k,l},\quad b_k=\sqrt{w_k}E_k^{\mathrm{To\text{-}fit}},
\end{equation}
Here, \(F_{k,l}\) is the precomputed contribution of basis function \(l\) to data point \(k\), and \(\mathbf c\) collects the corresponding linear coefficients, so \(\sum_l F_{k,l}c_l\) is the fitted exchange--correlation contribution to that data point.
We employ best-subset selection with mixed-integer optimization (MIO) to control sparsity~\cite{bertsimas2016best} using the Gurobi solver \cite{gurobi}:
\begin{equation}
\min_{\mathbf c,\mathbf z}\ \frac{1}{2}\|\mathbf A\mathbf c-\mathbf b\|^2\ \ \text{s.t.}\ \ \mathbf Q\mathbf c\le\mathbf t,\ -\mathcal M_U z_l\le c_l\le \mathcal M_U z_l,\ \sum_l z_l\le s,\ z_l \in \{0, 1\}.
\end{equation}
We scan $s=24$--80 and enforce numerical constraints (e.g., boundedness, one-electron/self-correlation, and Lieb--Oxford-type bounds).
A grid-sensitivity constraint restricts energy differences between (250,974)/SG-1 and (99,590)/SG-1 grids to $\le 0.015$ kcal/mol for selected high-sensitivity cases.
Full constraint definitions, mesh strategies, and solver settings are provided in the SI.

\paragraph{Design experiments, nonlinear parameters, and final choice.}
We performed a structured exploration of functional-design options, including: exchange non-uniform scaling correction (on/off), correlation base functional (SCAN vs PW92), same-spin self-correlation correction (on/off), $\tau$-indicator choices ($\beta$ vs $w$) for exchange, same-spin, and opposite-spin terms; polynomial-family choices (linear/Legendre/Chebyshev) for both $u$ and $v$ ($\beta$/$w$) expansions, and numerical-constraint settings.
To reduce cost while preserving coverage, we used an orthogonal $L_{36}$ design and evaluated candidates by the mean normalized error ratio (NER) over Gold-Standard Chemical Database (GSCDB).
We then refined key factors with full factorial tests and assessed the effect of numerical-constraint sets and grid-sensitivity thresholds.
We optimized nonlinear parameters that affect the functional form and dispersion by finite-grid scans: the range-separation parameter $\omega$, the VV10 damping parameter $b$ (with $C$ fixed), and the same-spin gradient nonlinearity $\gamma_{c,ss}$, including tests of alternative finite-domain transformations~\cite{MatitoWATOC2025}.
The final values are $\omega=0.27$, $b=5.5$, and $\gamma_{c,ss}=0.01$, while $C=0.01$, $\gamma_x=0.004$, and $\gamma_{c,os}=0.006$ are retained from prior parameterizations.
The full factor list, thresholds, nonlinear-parameter scans, and final fitting weights are provided in the SI.

\backmatter

\bmhead{Supplementary information}

\begin{itemize}
\item \textbf{SI.pdf}: Complete derivations, detailed assessments of constraint satisfaction, numerical procedures, design experiments, and comprehensive data and weighting tables.
\item \textbf{raw\_data.csv} and \textbf{statistics.xlsx}: Performance of the COACH functional on all benchmark sets.
\end{itemize}

\bmhead{Data availability}

The data supporting the findings of this study are available in the Supplementary Information and the accompanying \textbf{raw\_data.csv} and \textbf{statistics.xlsx} files.
The XYZ geometries, reference values, and evaluation metrics for all benchmark sets are publicly available through GSCDB at \url{https://github.com/JiashuLiang/GSCDB}.

\bmhead{Code availability}

The full COACH training protocol and workflow are publicly available at \url{https://github.com/JiashuLiang/COACH} and can be used to develop and optimize other density functionals.
A standalone PySCF implementation, example geometries, Q-Chem reference outputs, and regression tests for COACH are available at \url{https://github.com/JiashuLiang/COACH/tree/main/FunctionalCOACH}.
COACH is implemented as \texttt{XC\_HYB\_MGGA\_XC\_COACH} (ID 771) in the Libxc development branch and can be exposed by downstream programs once they support a Libxc version containing this functional.

\bmhead{Acknowledgements}

This work was supported by the Director, Office of Science, Office of Basic Energy Sciences, of the U.S. Department of Energy through the Gas Phase Chemical Physics Program, under Contract No. DE-AC02-05CH11231.
We thank Gurobi Optimization, LLC for providing the Gurobi Academic License used in this work.

We respectfully honor the memory of Axel D. Becke, whose foundational contributions to molecular density-functional theory include the multicenter partitioning scheme for numerical integration, the pioneering development of hybrid DFT, and the three-parameter hybrid framework that gave rise to B3LYP.
His influence continues to be felt throughout modern functional development, especially in the B97 family and the many semiempirical functionals derived from it, including the present work, for which we remain deeply grateful.

\section*{ Conflict of interest}
M.H-G. is a part owner of Q-Chem Inc., whose software was used in the calculations reported here.

\bibliography{sn-bibliography}

\end{document}